\documentclass[conference]{IEEEtran}
\IEEEoverridecommandlockouts
\usepackage{cite}
\usepackage{amsmath,amssymb,amsfonts}
\usepackage{algorithmic}
\usepackage{graphicx}
\usepackage{textcomp}
\usepackage{xcolor}
\def\BibTeX{{\rm B\kern-.05em{\sc i\kern-.025em b}\kern-.08em
    T\kern-.1667em\lower.7ex\hbox{E}\kern-.125emX}}
\begin{document}

\title{MaskPE: One Mask, Two Roles: Weight Placement and MAC Gating in Systolic Arrays \\
}


\author{
\IEEEauthorblockN{Shu-Lin Shih, Chandra Sekhar Mummidi, Aamir Gujral Elhaissouni, Sandip Kundu}
\IEEEauthorblockA{
University of Massachusetts Amherst, MA, USA\\
\vspace{-10mm}
}
}

\maketitle

\begin{abstract}
    Unstructured pruning reduces weight storage but does not inherently reduce execution cost in conventional weight-stationary systolic arrays. Packed nonzero weights must be restored to fixed processing-element positions, while zero-weight positions must preserve data forwarding. This paper presents MaskPE, a mask-directed systolic architecture that uses positional masks to steer nonzero weights during preload and clock-gate arithmetic registers in zero-weight processing elements during execution, reducing dynamic power while preserving systolic dataflow. At 50\% sparsity, MaskPE reduces compressed weight storage by 34.9\%, including metadata overhead. 
    RTL validation confirms exact execution with no additional steady-state tile stalls, while gate-level simulations demonstrate 12.5--14.4\% lower execution energy across samples of seven LLM projections.
\end{abstract}


\section{Introduction}
\label{sec:intro}

Systolic arrays achieve high throughput through regular data
movement and local operand reuse~\cite{Jouppi2017IndatacenterPA}.
In weight-stationary (WS) arrays, weights remain in processing
elements (PEs) while activations and partial sums traverse fixed
communication paths, providing predictable execution without
irregular scheduling.

Unstructured pruning challenges this regularity. Sparse
accelerators such as EIE~\cite{Han2016EIEEI},
SCNN~\cite{Parashar2017SCNNAA}, and Eyeriss
v2~\cite{Chen2018EyerissVA} exploit packed nonzeros and
positional metadata through indexing, operand selection, or
specialized scheduling~\cite{Zhang2016CambriconXAA,
Lu2021SangerAC,Zhao2024ChatOPUAF,Wang2025DenSparSAAB}.
A conventional WS array faces a different problem. 
Each compressed column contains variable numbers of nonzeros,
yet every PE requires its original weight within a fixed
preload schedule. Exploiting compressed residency therefore
requires restoring weight positions without disrupting
regular systolic execution.

\begin{figure}[t]
    \centering
    \includegraphics[width=\columnwidth]
    {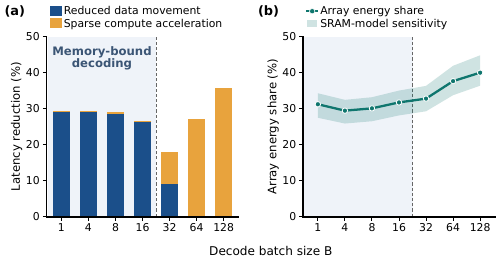}
    \vspace{-1em}
    \caption{Performance and energy implications of sparse
    weights in Llama~3.2~1B decoding.
    (a) Sparseloop-modeled latency reduction from reduced
    weight movement and sparse execution.
    (b) Systolic-array share of modeled Gemmini accelerator \cite{Gen2019GemminiES}
    energy, with SRAM-model sensitivity.
Shading marks memory-bound decoding ($M=1$--$16$).
    }
    \label{fig:motivation}
    \vspace{-5mm}
\end{figure}

This distinction is particularly important for low-batch large language model (LLM) decoding, where limited weight reuse makes data movement a major performance bottleneck. In practical serving, KV-cache capacity further constrains decode batch sizes, limiting the degree to which batching can improve compute utilization~\cite{Patel2023SplitwiseEG}. Fig.~\ref{fig:motivation}(a) shows that reducing weight movement lowers modeled latency by 26.6--29.4\% at batch sizes of 1--16, whereas accelerating sparse arithmetic provides little additional benefit. However, the systolic array still accounts for approximately 30\% of modeled accelerator energy under these conditions (Fig.~\ref{fig:motivation}(b)). Therefore, reducing weight movement improves performance, while suppressing unnecessary array activity remains important for energy efficiency.

\textbf{Architectural insight}:
Weight stationarity transforms irregular sparse-operand
matching into a once-per-tile placement problem. A weight's
original position needs to be resolved only during preload,
before activation processing begins. The next compressed
tile can be unpacked while the current one computes, provided 
variable-length weight streams are retrieved and positioned by the
usual preload deadline. Once placed, the stationary weight's zero/nonzero
status also provides a tile-invariant signal for suppressing
arithmetic activity without interrupting activation or
partial-sum forwarding.

We propose \emph{MaskPE}, a bitmap-directed WS systolic
architecture that exploits this insight. Packed nonzero
weights and positional metadata remain compressed throughout
storage and delivery. Banked storage, per-column
streaming decoders, and overlapped prefetch reconstruct
weights directly into their original PE positions at the
array boundary, without any intermediate weight buffer.
The conventional PE mapping, forwarding network, and
execution schedule remain unchanged.

MaskPE reuses the same positional bitmap for
two architectural roles: \textbf{weight placement,} directing packed weights to their
original positions during preload and \textbf{MAC gating} controlling
product-register clock gating during execution. Zero-weight
PEs suppress unnecessary arithmetic-register updates while
continuing activation and partial-sum forwarding. This
integrates compressed-weight residency and sparse-aware
activity control without in-array operand matching
or scheduling.

Our main contributions are:
\begin{itemize}
\item \textbf{Position-preserving compressed residency}:
    We introduce a WS architecture that retains packed
    nonzeros and positional metadata throughout on-chip
    storage and delivery, reconstructing weights directly
    into their original PE positions without dense staging
    or irregular sparse execution.

    \item \textbf{Deadline-aware delivery and unified
    bitmap control}:
    A banked delivery engine with per-column streaming
    decoders and overlapped prefetch satisfies the
    conventional preload schedule without additional
    steady-state stalls. The same bitmap controls weight
    placement and product-register gating while preserving
    activation and partial-sum forwarding.

    \item \textbf{Register-transfer level (RTL) implementation and architectural
    evaluation}:
    We implement a 32 × 32 INT8/INT32 MaskPE array with RTL compressed-weight delivery and streaming decoders and validate exact execution on a Wanda-pruned~\cite{sun2024simple} Llama 3.2 1B projection. 
At 50\% sparsity, MaskPE reduces weight storage by
34.9\%, logic-plus-weight-SRAM area by 15.04\%,
and post-CTS logic-plus-modeled-weight-memory energy
by 12.5--14.4\% across samples of seven LLM projections.
\end{itemize}

\noindent\textbf{Novelty:} Prior work uses sparsity metadata to direct nonzero operands to compute units~\cite{Han2016EIEEI,Parashar2017SCNNAA,Chen2018EyerissVA,Lu2021SangerAC}, or uses zero weights to disable computation while still delivering weights through a dense data path~\cite{Pandey2021UPTPUIE,Xue2025ReGateEP}. MaskPE uses a single positional mask for both purposes. During preload, the bitmap identifies the locations of nonzero weights in the compressed weight stream. During execution, the same bitmap serves as the clock-enable signal for arithmetic registers. This approach eliminates the need for an \textit{index-matching network, load balancer, or separate gating predictor}. It also allows the same metadata to support both storage reduction and power savings, avoiding the overhead of maintaining separate metadata for each purpose. Since nonzero weights remain assigned to the same PEs, the activation and partial-sum schedules remain unchanged from the dense design. To our knowledge, this is the first design to demonstrate at both RTL and gate level that a single mask can serve as both a placement directive and a gating signal, making unstructured sparsity beneficial for a conventional weight-stationary array.

\section{Background and Related Work}
\label{sec:background}

\subsection{Structured and Unstructured Sparsity}
\label{sec:sparsity_quality}

\begin{figure}[t]
    \centering
    \includegraphics[width=0.6\columnwidth]{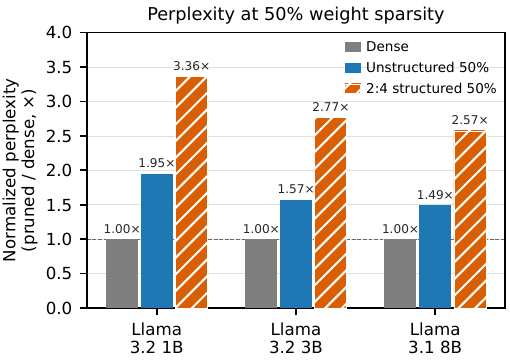}
    \caption{Perplexity relative to dense models at 50\%
    sparsity for Llama~3.2~1B, Llama~3.2~3B, and
    Llama~3.1~8B.}
    \label{fig:pruning_quality}
    \vspace{-5mm}
\end{figure}

Unstructured pruning removes individual weights without
constraining their positions~\cite{LeCun1989OptimalBD,Frantar2023SparseGPTML},
whereas structured sparsity restricts removal patterns to
simplify hardware execution. At 50\% sparsity, unstructured
pruning increases perplexity by 1.49--1.95$\times$ across
three Llama models, compared with 2.57--3.36$\times$
for 2:4 pruning (Fig.~\ref{fig:pruning_quality}).
This quality gap motivates hardware that supports arbitrary
weight sparsity without imposing additional pruning constraints.

\subsection{Weight-Stationary Systolic Execution}
\label{sec:ws}

Weight-stationary systolic arrays exploit local reuse
and regular data movement~\cite{Jouppi2017IndatacenterPA}.
Each processing element  holds a stationary weight
while activations and partial sums traverse fixed paths.
For weight $w_{i,j}$, the PE computes
\begin{equation}
p_{\mathrm{out}}=p_{\mathrm{in}}+a_iw_{i,j}.
\label{eq:pe_mac}
\end{equation}
When $w_{i,j}=0$, arithmetic is unnecessary, but activation
and partial-sum forwarding must continue to preserve the
systolic schedule. Weight stationarity makes the zero/nonzero
status invariant throughout tile execution, enabling
selective arithmetic activity control without changing
operand routing.

\subsection{Packed Unstructured Sparsity}
\label{sec:packed_sparse}

Compressed sparse architectures such as
EIE~\cite{Han2016EIEEI}, SCNN~\cite{Parashar2017SCNNAA},
and Eyeriss v2~\cite{Chen2018EyerissVA} reduce storage
and movement by retaining nonzero operands with positional
metadata. Exploiting these representations during computation
requires associating operands with their corresponding
activations and accumulation destinations.

For a tile containing $L$ weight positions, a bitmap
$M\in\{0,1\}^{L}$ identifies nonzeros stored consecutively
in a value stream $V$. The index of a retained value is
\begin{equation}
r(i)=\sum_{k=0}^{i-1}M_k,\qquad M_i=1.
\label{eq:bitmap_rank}
\end{equation}
With $b$-bit weights and sparsity $s$, this representation
requires ideally $L[b(1-s)+1]$ bits versus $Lb$ dense bits,
excluding alignment and buffering overhead.

Unlike sparse execution architectures that dynamically
associate operands, a fixed-position WS array requires
each packed weight to reach its original PE. The central
challenge is therefore to recover positional information
without forfeiting compressed-data movement or disrupting
the regular execution schedule.

\subsection{Sparse Arithmetic and Clock Gating}
\label{sec:gating}

Clock gating suppresses register updates without changing
the underlying dataflow. Given a stationary positional bit
$M_{i,j}$ and activation-valid signal $v_{i,j}(t)$,
the product-register enable is
\begin{equation}
e_{i,j}(t)=v_{i,j}(t)\land M_{i,j}.
\label{eq:mac_enable}
\end{equation}
Zero-weight PEs can thus suppress product-register activity
while preserving operand forwarding. The energy benefit
depends on clock-tree savings relative to gating overhead,
residual combinational switching, and leakage.

Prior work explores low-power systolic execution through
data encoding and reordering~\cite{Peltekis2023ExploitingDE},
multiply-accumulate (MAC) power gating with bypass forwarding~\cite{Pandey2021UPTPUIE},
and fine-grained power management~\cite{Xue2025ReGateEP}.
These mechanisms establish opportunities for reducing
arithmetic activity, distinct from reducing weight movement.

\subsection{Sparse Systolic Execution and Prior Work}
\label{sec:prior_sparse}

Sparse accelerators exploit nonzero operands through
indexing~\cite{Zhang2016CambriconXAA}, constrained
sparsity patterns~\cite{Liu2021S2TAES}, or specialized
execution: Sanger~\cite{Lu2021SangerAC} uses reconfigurable
execution for sparse attention, ChatOPU~\cite{Zhao2024ChatOPUAF}
uses compressed weights in a diagonal systolic array, and
DenSparSA~\cite{Wang2025DenSparSAAB} supports both dense
and unstructured sparse execution.

These approaches illustrate the hardware and scheduling
tradeoffs of sparse computation. MaskPE instead targets
compressed weight residency and selective arithmetic
activity while retaining fixed PE assignments and conventional
WS execution. Its architectural tradeoff is the cost of
position-preserving weight delivery against the storage,
movement, and switching activity it eliminates.

\section{MaskPE Architecture}
\label{sec:darkpe_architecture}

\subsection{Architecture Overview and Tile Execution}
\label{subsec:arch}

\begin{figure}[t]
    \centering
    \includegraphics[width=0.7\columnwidth]{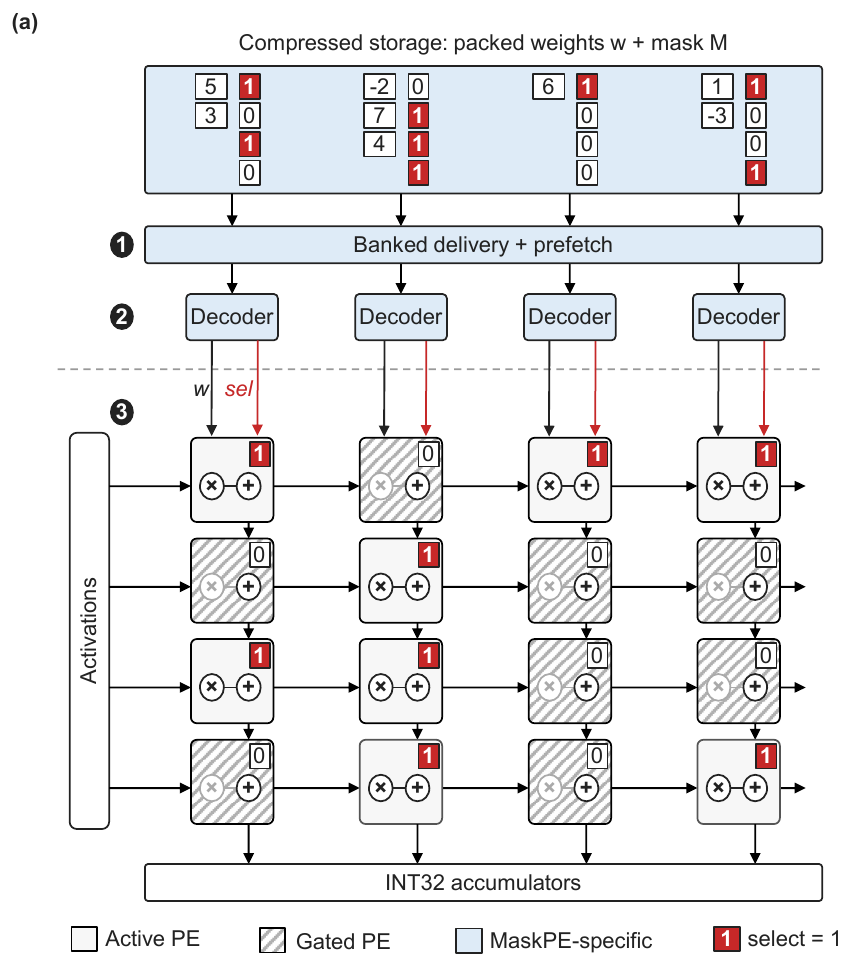}
    \includegraphics[width=0.7\columnwidth]{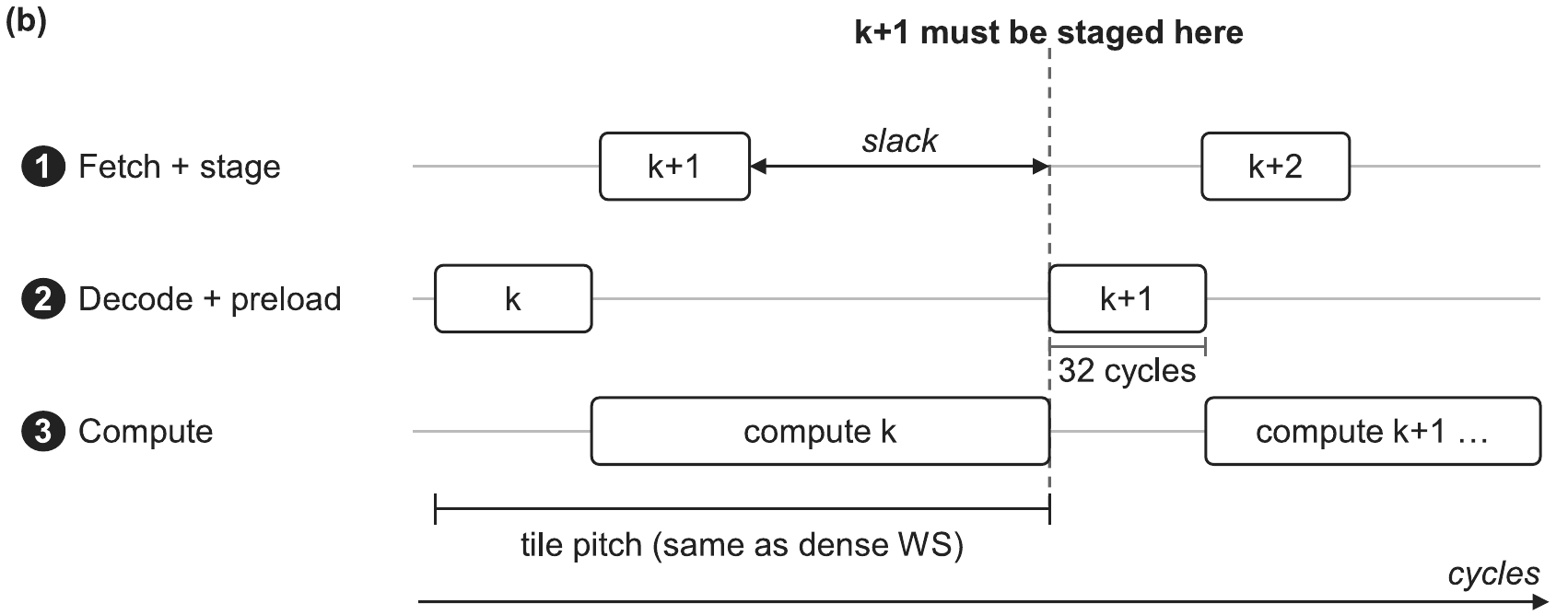}
    \caption{MaskPE architecture. (a) Compressed-weight
    delivery, position-aligned preload, and regular WS
    execution. \textbf{The same positional bitmap directs weight placement during preload and MAC gating during execution.} (b) Next-tile weight staging overlaps
    current-tile computation, meeting the fixed preload 
    deadline without extending the tile interval.}
    \label{fig:tile_org}
    \vspace{-7mm}
\end{figure}

MaskPE separates compressed-weight preparation from
regular weight-stationary execution
(Fig.~\ref{fig:tile_org}(a)).
During computation of tile~$k$, banked delivery fetches
and stages the packed weights and positional masks for
tile~$k{+}1$ (\textcircled{\scriptsize 1}). During the fixed 32-cycle preload,
the column decoders use these masks to reconstruct
position-aligned weight and selection-bit pairs
$(w,\mathrm{sel})$ (\textcircled{\scriptsize 2}). Each PE retains its assigned
weight and selection bit. During computation, the
selection bit gates the product register at zero-weight
positions, while activation and partial-sum forwarding
follow the conventional WS dataflow (\textcircled{\scriptsize 3}).

Weight staging overlaps computation of the preceding
tile (Fig.~\ref{fig:tile_org}(b)). When staging completes
before the next preload deadline, compressed-weight
delivery adds no steady-state stalls, preserving the
conventional tile pitch. Successive array computations
do not overlap. Outputs accumulate in 32 column-aligned
INT32 banks without sparsity-dependent routing because
each PE retains its original weight assignment.

\begin{figure*}[t]
    \centering
    \includegraphics[width=0.8\linewidth]{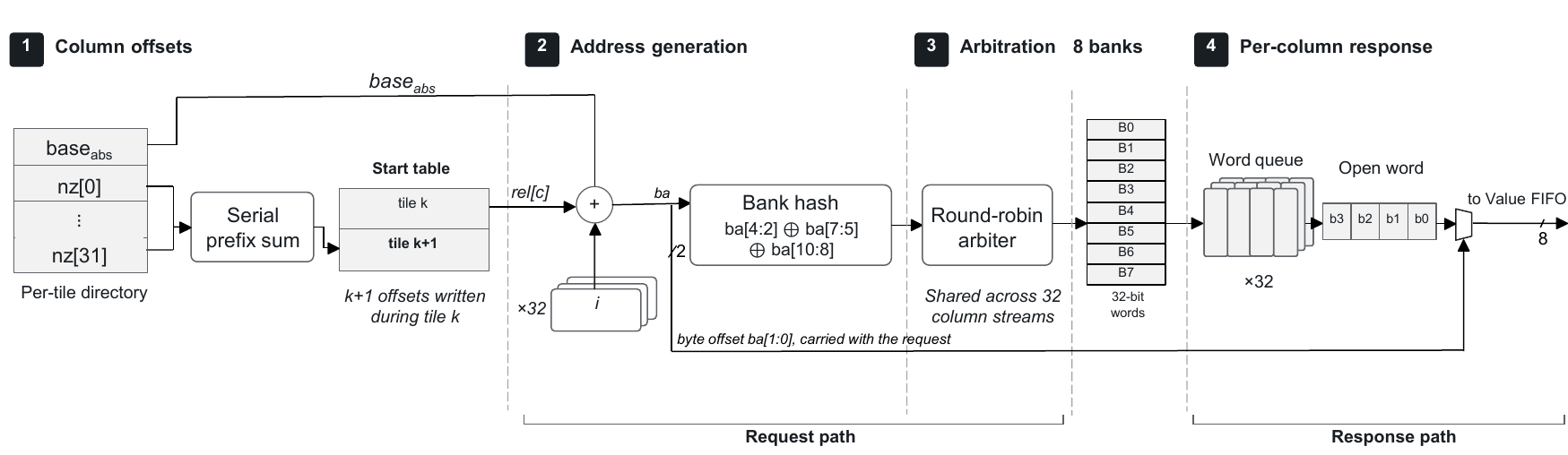}
    \caption{MaskPE compressed-weight delivery: column-offset
    generation, banked requests, and per-column response buffering.}
    \label{fig:weight_delivery}
\end{figure*}

\subsection{Position-Preserving Sparse Execution}
\label{subsec:position_preserving}

\begin{figure}[t]
    \centering
    \includegraphics[width=0.75\columnwidth]{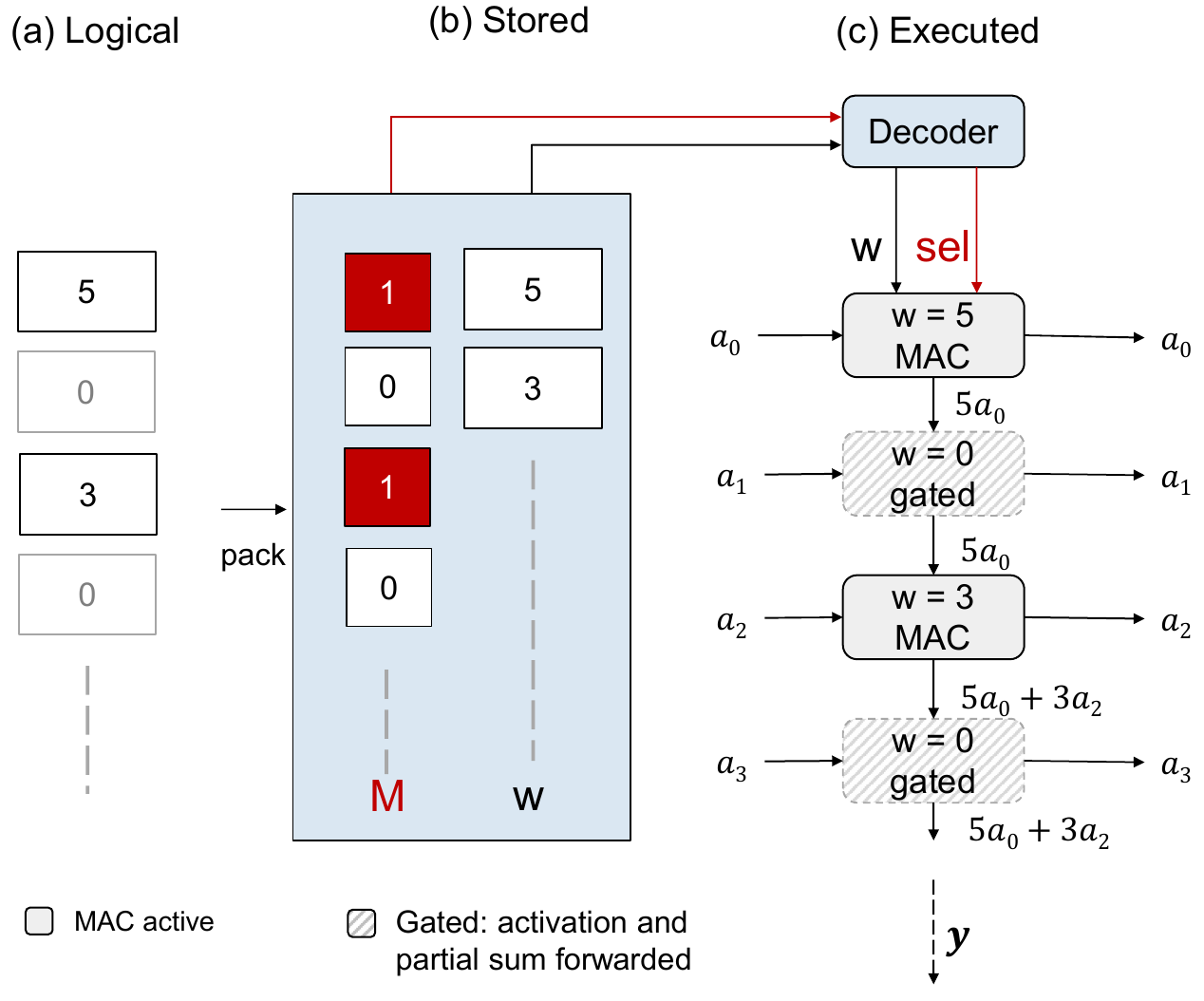}
\caption{Bitmap-directed reconstruction restores packed weights to their original PE positions while gating product registers at zero-weight PEs and preserving forwarding.}
    \label{fig:position_preserving}
    \vspace{-7mm}
\end{figure}

Packed nonzeros cannot be loaded consecutively into a WS
array without changing their assigned computations.
For example, loading the packed column $[5,3]$ directly
instead of its original representation $[5,0,3,0]$
would produce $5a_0+3a_1$ rather than
$5a_0+3a_2$ (Fig.~\ref{fig:position_preserving}).

MaskPE resolves this positional mismatch during preload.
Each column decoder consumes packed values according to
the bitmap, emitting nonzero weights at their original
positions and zeros elsewhere. The resulting preload
stream preserves the conventional PE mapping without
an intermediate dense weight buffer or irregular in-array
operand matching.

The same bitmap supplies a selection bit stored alongside
each stationary weight. During execution, this bit gates
the product register at zero-weight PEs, while activation
and partial-sum forwarding continue unchanged. Thus,
a single positional representation supports both weight
placement and selective arithmetic activity control.

\subsection{Compressed-Weight Organization}
\label{subsec:compressed_weights}

Unstructured sparsity produces variable nonzero counts
across columns, making fixed-size column regions and
word-aligned streams inefficient. MaskPE therefore packs
nonzero INT8 weights consecutively across columns and
tiles without inter-column padding, storing positional
bitmaps separately.

A per-tile directory records the packed-value base address
and nonzero count of each column. Prefix sums recover
column offsets without storing individual absolute
addresses, allowing streams to begin at arbitrary byte
positions within memory words.

Although compressed columns have variable lengths,
their preload schedules remain fixed.
Consequently, the delivery engine must supply every
nonzero before its preload slot. MaskPE
satisfies this deadline through overlapped prefetch
and per-column staging, described in Section~IV.

\section{MaskPE Microarchitecture}
\label{sec:microarchitecture}

\subsection{Banked Storage and Compressed-Weight Delivery}
\label{sec:delivery}

Figure~\ref{fig:weight_delivery} details the banked delivery
and prefetch stage of Fig.~\ref{fig:tile_org}(a). Byte-tight
packing eliminates padding between compressed columns but
produces variable-length streams with unaligned starting
addresses. MaskPE generates addresses for these streams,
arbitrates their accesses across eight single-read-port banks,
and buffers returned values independently for each column.
This organization separates irregular memory access from
the fixed-rate weight preload of the systolic array.

\noindent\textbf{1) Column offsets:}
A 27-byte per-tile directory stores a 24-bit packed-value
base address $\mathrm{base}_{\mathrm{abs}}$ and 32 six-bit
nonzero counts. A serial prefix-sum generator converts
these counts into the starting offset $\mathrm{rel}[c]$
of each compressed column. The offsets are held in a
two-slot start table, allowing the current tile's addresses
to remain available while the next tile's offsets are
generated. As part of the overlapped preparation in
Fig.~\ref{fig:tile_org}(b), offsets for tile~$k{+}1$
are computed during tile~$k$. Prefix-sum generation
therefore lies outside the next tile's weight-request
critical path.

\noindent\textbf{2) Address generation:}
For nonzero index $i$ in column $c$, the byte address is
\begin{equation}
\mathrm{ba}=\mathrm{base}_{\mathrm{abs}}+\mathrm{rel}[c]+i.
\end{equation}
Each column advances its own index $i$ because unstructured
sparsity produces different nonzero counts across columns.
The resulting byte addresses identify both the memory
locations of packed values and their positions within
32-bit words. This permits consecutive nonzeros to be
stored without column-alignment padding.

\noindent\textbf{3) Arbitration across eight banks:}
The 32 column streams may request weights concurrently,
but the eight memory banks each provide one read port.
Bank selection uses an XOR of address fields to distribute
requests:
\begin{equation}
\mathrm{bank}=\mathrm{ba}[4\!:\!2]\oplus
\mathrm{ba}[7\!:\!5]\oplus\mathrm{ba}[10\!:\!8].
\end{equation}
A shared round-robin arbiter schedules competing requests
across the available banks. Arbitration allows variable
column demands to share the memory ports without
requiring a dedicated read port for every decoder.

\noindent\textbf{4) Per-column response:}
Each bank returns 32-bit words to per-column word queues.
The byte offset $\mathrm{ba}[1\!:\!0]$, carried with the
request, selects the required INT8 value from the returned
word. The extracted values are supplied in order to
the corresponding decoder FIFOs. Per-column buffering
absorbs variations in bank service and decouples
compressed-weight retrieval from the fixed preload
sequence.

\noindent\textbf{Overlapped prefetch:}
Because compressed columns have variable nonzero counts
and share the banked memory system, their retrieval
times need not be identical. MaskPE fetches and stages
tile~$k{+}1$ during computation of tile~$k$, as shown
in Fig.~\ref{fig:tile_org}(b). The staged values are
then consumed by the column decoders during the next
tile's fixed 32-cycle preload. Completing staging before
that preload begins prevents irregular bank service
from introducing additional steady-state stalls.

\subsection{Position-Preserving Streaming Decoder}
\label{sec:decoder}

\begin{figure}[t]
    \centering
    \includegraphics[width=0.8\columnwidth]{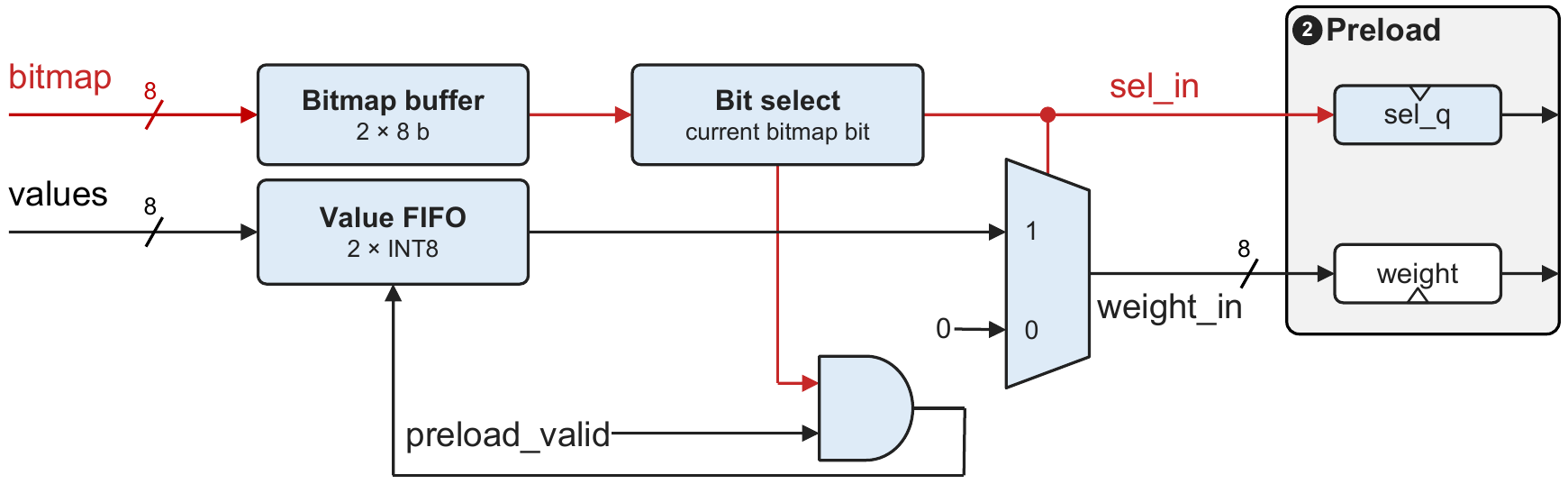}
    \caption{Position-preserving streaming decoder.}
    \label{fig:decoder_pe}
    \vspace{-5mm}
\end{figure}

Each array column has a streaming decoder with a
two-entry bitmap-byte buffer and a two-entry INT8 value
first-in, first-out (FIFO) (Fig.~\ref{fig:decoder_pe}). The decoder traverses
rows 31 to 0 in bottom-first preload order, processing
bitmap bits least significant bit (LSB)-first.

For each set bit, the decoder consumes the next packed
value and emits it with $\mathit{sel}=1$; for each
cleared bit, it emits zero with $\mathit{sel}=0$
without advancing the value FIFO. Thus, every column
generates exactly 32 position-aligned preload outputs
regardless of its nonzero count, without reconstructing
a dense column in local storage.

The receiving PE captures the weight and selection bit.
A missing value at a set-bit position raises an underflow
flag rather than stalling the decoder, making timely
prefetch essential to correct execution.

\subsection{Sparse-Aware Processing Element}
\label{sec:pe}

\begin{figure}[t]
    \centering
    \includegraphics[width=0.8\columnwidth]{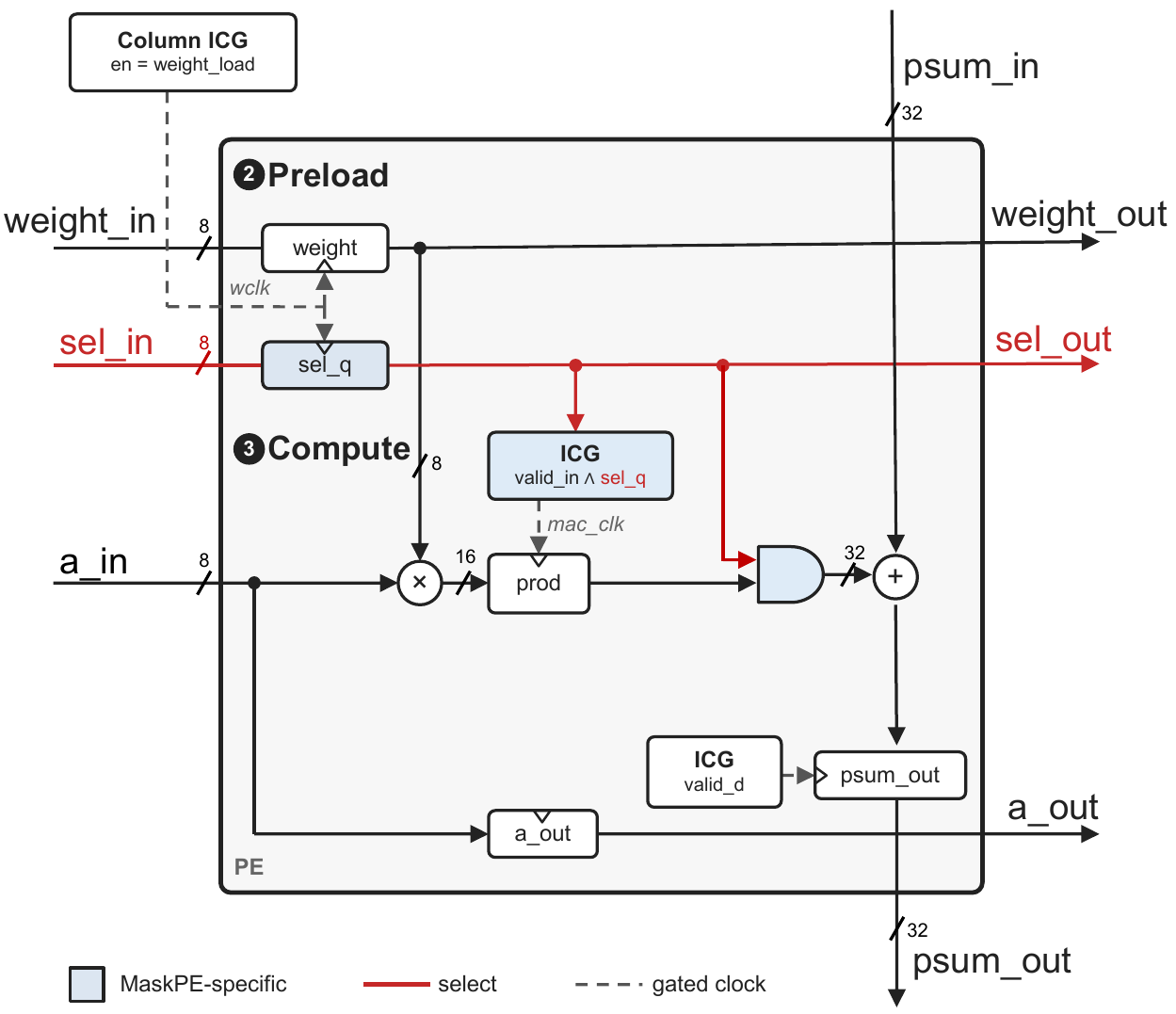}
    \caption{MaskPE processing element. During preload
    \textcircled{\scriptsize 2}, a column-shared clock gate loads the stationary
    weight and bitmap select bit. During computation
    \textcircled{\scriptsize 3}, the select bit gates the product register and
    masks its output before partial-sum accumulation.}
    \label{fig:pe}
    \vspace{-5mm}
\end{figure}

Fig.~\ref{fig:pe} shows how MaskPE incorporates sparsity
control into a conventional weight-stationary (WS) PE.
Following compressed-weight delivery, step \textcircled{\scriptsize 2} loads the
weight and its bitmap-derived select bit $\mathit{sel}_q$
through a column-shared preload clock gate. Both values
propagate down the column during preload and remain
stationary during computation.

In step \textcircled{\scriptsize 3}, the PE retains the regular activation and
partial-sum forwarding paths. The product register is
clock-gated with
\begin{equation}
\mathit{en}_{\mathrm{MAC}} =
\mathit{valid}_{\mathrm{in}} \land \mathit{sel}_q,
\end{equation}
suppressing register updates at pruned positions. Since
the gated register may retain a stale product, its
sign-extended output is masked before accumulation:
\begin{equation}
\begin{aligned}
\mathit{prod}_{\mathrm{eff}}
    &= \operatorname{sext}_{32}(\mathit{prod})
       \land \{32\{\mathit{sel}_q\}\},\\
\mathit{psum}_{\mathrm{next}}
    &= \mathit{psum}_{\mathrm{in}}
       + \mathit{prod}_{\mathrm{eff}}.
\end{aligned}
\end{equation}

The 32-bit accumulation uses a conventional adder,
with the product masked before accumulation rather
than bypassing the adder output. The partial-sum output
register is independently clock-gated by
$\mathit{valid}_d$, while activation forwarding remains
ungated. Thus, bitmap-controlled product suppression
preserves the WS dataflow and fixed execution schedule.

\section{Experiments}
\label{sec:exps}


\subsection{Experimental Methodology}
\label{sec:eval:method}

\noindent\textbf{Architecture and workload}:
We implement MaskPE as a $32\times32$ weight-stationary
systolic array with INT8 multiplication and INT32 accumulation. We evaluate seven layer-8 Llama~3.2~1B projections
at 50\% Wanda~\cite{sun2024simple} unstructured sparsity with INT8 weights.
Energy evaluation samples three output tiles per projection;
the remaining experiments use a 512-tile
\texttt{gate\_proj} slice ($K=2048$, $N=256$)
at $M=128$, with $M=1,4,16$ for batch sensitivity.

Identical weights and input activations are used across
all compared architectures. RTL correctness, physical area, the detailed energy breakdown, and bandwidth evaluations use this projection workload.


\smallskip
\noindent\textbf{Architectural comparisons}:
We compare MaskPE against a conventional dense
weight-stationary systolic array with identical PE
dimensions, arithmetic precision, and execution schedule.
Both use conventional INT32 accumulation without the
additional generic PE optimizations considered during
design exploration. MaskPE adds compressed-weight
delivery and bitmap-controlled product-register gating
and masking. The conventional dense WS implementation
serves as the reference for area, energy, and performance.

\smallskip
\noindent\textbf{Hardware implementation and energy analysis}:
All architectures are described in RTL and synthesized
using Yosys/ABC~\cite{Wolf2013YosysAFV} with the Nangate45 standard-cell library.
Functional correctness and execution cycles are verified
against the same dense reference.
For energy evaluation, gate-level switching activity is
collected from the frozen workload and analyzed with
OpenSTA~\cite{CherryOpenSTA}. Placement and clock-tree synthesis are performed
using OpenROAD~\cite{Ajayi2019OpenROADTA}, with placed parasitic estimates used
for post-CTS logic-energy analysis.
Weight-SRAM area uses FakeRAM45 memory macros~\cite{TILOS_NanGate45}, while memory energy is estimated from recorded accesses using Accelergy \cite{Wu2019AccelergyAA} energy coefficients. Combined energy includes logic and modeled weight-memory energy, excluding common activation and accumulator memories.

\smallskip
\noindent\textbf{Bandwidth and timing methodology}:
To characterize the interaction between compressed-weight
delivery and systolic execution, we evaluate bandwidth
budgets of 2, 4, 8, 16, and 32 bytes/cycle per array.
An external tile-granular memory-service model controls
weight availability through the existing tile-start
handshake, while the RTL determines execution cycles
and verifies computation. The experiment therefore
captures bandwidth-induced tile delays without modeling
individual dynamic random-access memory (DRAM) transactions or cycle-level backpressure.
Physical evaluation uses a 2.5\,ns target clock period
and post-CTS placement-estimated parasitics. Timing
checks use the documented 0.8\,ns host-interface
clock-insertion assumption.

\smallskip
\noindent\textit{B: Position-Preserving Execution}:
MaskPE preserves conventional weight-stationary execution
despite variable-length compressed weight streams.
Across 512 evaluated weight tiles, all 16,384 decoded
column streams match the dense reference, and all
32,768 outputs are exact. The tile pitch remains
identical to conventional WS execution:
100, 103, 115, and 227 cycles for $M=1,4,16,128$,
respectively. Next-tile preparation completes in
approximately 36 cycles, overlapping computation
without additional steady-state stalls. Correctness
and schedule preservation also hold across directed
sparsity patterns and synchronous SRAM interfaces.

\begin{figure}[t]
    \centering
    \includegraphics[width=\columnwidth]{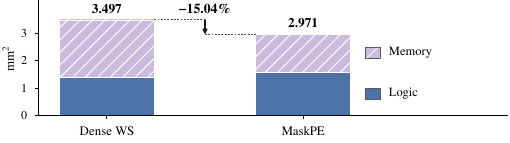}
    \caption{Area cost of compressed-weight residency.
    Logic and FakeRAM45-modeled weight-SRAM area.}
    \label{fig:vc_area}
    \vspace{-5mm}
\end{figure}

\smallskip
\noindent\textit{C. Compressed Residency and Area Efficiency}:
Preserving the conventional WS schedule while reconstructing
compressed weights requires additional delivery and control
logic. However, keeping weights compressed throughout on-chip
storage reduces the SRAM capacity required by the array.
Fig.~\ref{fig:vc_area} quantifies this tradeoff: the area
saved through compressed weight residency exceeds the
additional logic cost, reducing total logic-plus-weight-SRAM
area by 15.04\% relative to conventional dense WS.
Thus, MaskPE achieves position-preserving sparse execution
without sacrificing area benefits.

\begin{figure}[t]
    \centering
    \includegraphics[width=\columnwidth]{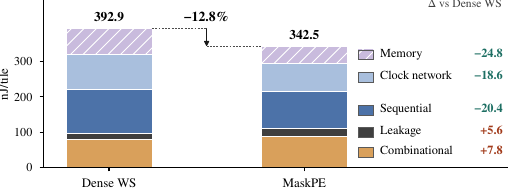}
    \caption{Post-CTS energy breakdown for the frozen
\texttt{gate\_proj} workload: conventional dense WS
versus MaskPE.}

    \label{fig:vd_energy}
    \vspace{-5mm}
\end{figure}

\smallskip
\noindent\textit{D. Physical-Aware Energy Breakdown}:
MaskPE offsets compressed-weight reconstruction costs
through reduced weight movement and clocked activity.
Across seven Llama~3.2~1B projections, post-CTS
hybrid-energy savings range from 12.5\% to 14.4\%,
with a 13.2\% weight-tile-count-weighted estimate
from sampled tiles. Fig.~\ref{fig:vd_energy}
details the frozen \texttt{gate\_proj} case:
sequential, clock-network, and modeled weight-memory
energy decrease, outweighing higher combinational
and leakage energy to yield a 12.8\% reduction.

\begin{figure}[t]
    \centering
    \includegraphics[width=0.8\columnwidth]{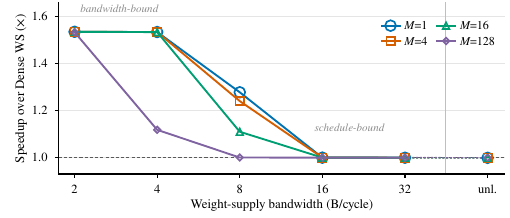}
    \caption{Bandwidth-dependent MaskPE speedup over
    conventional dense WS.}
    \label{fig:ve_bandwidth}
    \vspace{-5mm}
\end{figure}

\smallskip
\noindent\textit{E. Bandwidth-Dependent Performance}:
MaskPE preserves the conventional WS compute schedule,
so its performance advantage arises when weight delivery
limits execution. Fig.~\ref{fig:ve_bandwidth} shows that
compressed weight streams reduce delivery time under
bandwidth constraints, achieving up to $1.53\times$
speedup over dense WS. As weight-supply bandwidth
increases, both architectures become limited by the
same systolic schedule, and the speedup converges to
$1\times$. Larger batch sizes reach this regime earlier
because longer compute intervals increasingly hide
weight delivery. Thus, MaskPE converts compressed
residency into performance gains when data movement,
rather than computation, determines execution time.

\begin{figure}[t]
    \centering
    \includegraphics[width=0.8\columnwidth]{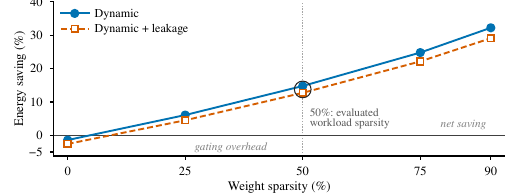}
    \caption{Sparsity-dependent energy savings of MaskPE
relative to conventional dense WS, including standard-cell
and modeled weight-memory energy.}
    \label{fig:vf_gating}
    \vspace{-5mm}
\end{figure}

\smallskip
\noindent\textit{F. Sparsity-Dependent Energy Savings}:
MaskPE reuses positional metadata to suppress product-register
activity at zero-weight PEs while preserving activation and
partial-sum forwarding. Fig.~\ref{fig:vf_gating} shows how
the overall energy benefit relative to conventional dense WS
depends on weight sparsity. At zero sparsity, the additional
sparse-weight storage and control overhead increases energy
by 2.6\%. As sparsity increases, reduced weight movement
and product-register activity outweigh these costs,
yielding a 12.7\% energy reduction at the evaluated 50\%
sparsity and 29.1\% at 90\% sparsity.

\smallskip
\noindent\textit{G. One Mask, Two Roles: Energy Attribution}:
MaskPE uses the positional bitmap first to reconstruct
compressed weights during preload and then to suppress
product-register activity during execution. To isolate the second role, we disable bitmap-controlled
clock gating while keeping the PE datapath unchanged.
At 50\% sparsity, bitmap reuse reduces pre-CTS PE-array
energy by 5.1\% without altering the systolic schedule.



\smallskip
\noindent\textit{H. Physical Implementation and Timing}:
We implement MaskPE in Nangate45 with a target frequency
of 400\,MHz (2.5\,ns clock period). Physical timing
analysis shows that weight reconstruction, compressed
delivery, and PE computation meet their setup constraints.
All
reported setup, hold, clock-gating, and recovery/removal
checks pass with zero violations under the documented
0.8\,ns host-interface clock-insertion assumption.

\section{Conclusion}
\label{sec:conclusion}

MaskPE enables unstructured weight sparsity in conventional
weight-stationary systolic arrays through bitmap-directed
weight placement and selective product-register clock gating.
Packed nonzero weights are restored to their original PE
positions, preserving regular activation and partial-sum
forwarding without introducing irregular computation
scheduling. RTL validation confirms exact execution with
no additional steady-state tile stalls. Post-CTS evaluation shows 15.0\% lower SRAM-inclusive
area and 12.5--14.4\% lower combined standard-cell
and modeled weight-memory energy across samples of
seven LLM projections at 50\% sparsity, relative
to conventional dense WS.
Under bandwidth-constrained execution, compressed-weight
delivery achieves up to $1.53\times$ speedup. These results
demonstrate that compressed-weight residency and
bitmap-controlled arithmetic activity can reduce storage,
energy, and weight-delivery costs while retaining the
regular execution structure of weight-stationary systolic
arrays.

\bibliographystyle{IEEEtran}
\bibliography{references}

@article{Jouppi2017IndatacenterPA,
  title={In-datacenter performance analysis of a tensor processing unit},
  author={Norman P. Jouppi and Cliff Young and Nishant Patil and David A. Patterson and Gaurav Agrawal and Raminder Singh Bajwa and Sarah F. Bates and Suresh Bhatia and Nanette J. Boden and Alwin Dipl-Ing Borchers and R. Richard Milwaukee Boyle and Pierre-luc Cantin and Clifford Chao and Chris Clark and Jeremy Coriell and Michael Daley and M.T. Dau and Jeffrey Dean and Ben Gelb and Taraneh Ghaemmaghami and Rajendra Gottipati and William Gulland and Robert B. Hagmann and C. Richard Ho and Doug Hogberg and John Hu and Robert Hundt and Dan Hurt and Julian Ibarz and Aaron Jaffey and Alek Jaworski and Alexander Kaplan and Harshit Khaitan and Daniel Killebrew and A. Koch and Naveen Kumar and Steve Lacy and James Laudon and James Law and Diem Duy Le and Chris Leary and Zhuyuan Liu and Kyle Lucke and Alan Lundin and Gordon MacKean and Adriana Maggiore and Maire Mahony and Kieran Miller and Rahul Nagarajan and Ravi Narayanaswami and Ray Ni and Kenneth Nix and Thomas Norrie and Mark Omernick and Narayana Penukonda and Andy Phelps and Jonathan Ross and Matt Ross and Amir Salek and Emad Samadiani and Chris Severn and Gregory Sizikov and Matthew Snelham and Jed Souter and Dan Steinberg and Andy Swing and Mercedes Tan and Gregory Michael Thorson and Boyu Tian and Horia Toma and Elena Del Giudice Tuttle and Vijay K. Vasudevan and Richard Paul Walter and Walter Wang and Eric Wilcox and Doe Hyun Yoon},
  journal={2017 ACM/IEEE 44th Annual International Symposium on Computer Architecture (ISCA)},
  year={2017},
  pages={1-12},
}

@article{Han2016EIEEI,
  title={EIE: Efficient Inference Engine on Compressed Deep Neural Network},
  author={Song Han and Xingyu Liu and Hui-Zi Mao and Jing Pu and Ardavan Pedram and Mark Horowitz and William J. Dally},
  journal={2016 ACM/IEEE 43rd Annual International Symposium on Computer Architecture (ISCA)},
  year={2016},
  pages={243-254},
}

@article{Parashar2017SCNNAA,
  title={SCNN: An accelerator for compressed-sparse convolutional neural networks},
  author={Angshuman Parashar and Minsoo Rhu and Anurag Mukkara and Antonio Puglielli and Rangharajan Venkatesan and Brucek Khailany and Joel S. Emer and Stephen W. Keckler and William J. Dally},
  journal={2017 ACM/IEEE 44th Annual International Symposium on Computer Architecture (ISCA)},
  year={2017},
  pages={27-40},
}

@article{Chen2018EyerissVA,
  title={Eyeriss v2: A Flexible Accelerator for Emerging Deep Neural Networks on Mobile Devices},
  author={Yu-Hsin Chen and Tien-Ju Yang and Joel S. Emer and Vivienne Sze},
  journal={IEEE Journal on Emerging and Selected Topics in Circuits and Systems},
  year={2018},
  volume={9},
  pages={292-308},
}

@article{Zhang2016CambriconXAA,
  title={Cambricon-X: An accelerator for sparse neural networks},
  author={Shi-Jin Zhang and Zi-Dong Du and Lei Zhang and Hui-Ying Lan and Shao-Li Liu and Ling Li and Qi Guo and Tian-Shi Chen and Yun-Ji Chen},
  journal={2016 49th Annual IEEE/ACM International Symposium on Microarchitecture (MICRO)},
  year={2016},
  pages={1-12},
}

@article{Liu2021S2TAES,
  title={S2TA: Exploiting Structured Sparsity for Energy-Efficient Mobile CNN Acceleration},
  author={Zhi-gang Liu and Paul N. Whatmough and Yu-Hao Zhu and Matthew Mattina},
  journal={2022 IEEE International Symposium on High-Performance Computer Architecture (HPCA)},
  year={2021},
  pages={573-586},
}

@article{Lu2021SangerAC,
  title={Sanger: A Co-Design Framework for Enabling Sparse Attention using Reconfigurable Architecture},
  author={Li-Qiang Lu and Yicheng Jin and Hangrui Bi and Zizhang Luo and Peng Li and Tao Wang and Yun Liang},
  journal={MICRO-54: 54th Annual IEEE/ACM International Symposium on Microarchitecture},
  year={2021},
}

@article{Zhao2024ChatOPUAF,
  title={ChatOPU: An FPGA-Based Overlay Processor for Large Language Models with Unstructured Sparsity},
  author={Tiandong Zhao and Shaoqiang Lu and Chen Wu and Lei He},
  journal={2024 ACM/IEEE International Conference On Computer Aided Design (ICCAD)},
  year={2024},
  pages={1-9},
}

@article{Wang2025DenSparSAAB,
  title={DenSparSA: A Balanced Systolic Array Approach for Dense and Sparse Matrix Multiplication},
  author={Zi-Heng Wang and Ruiqi Sun and Xin He and Tian-Rui Ma and An Zou},
  journal={2025 62nd ACM/IEEE Design Automation Conference (DAC)},
  year={2025},
  pages={1-7},
}

@article{Pandey2021UPTPUIE,
  title={UPTPU: Improving Energy Efficiency of a Tensor Processing Unit through Underutilization Based Power-Gating},
  author={Pramesh Pandey and Noel Daniel Gundi and Koushik Chakraborty and Sanghamitra Roy},
  journal={2021 58th ACM/IEEE Design Automation Conference (DAC)},
  year={2021},
  pages={325-330},
}

@article{Peltekis2023ExploitingDE,
  title={Exploiting data encoding and reordering for low-power streaming in systolic arrays},
  author={Christodoulos Peltekis and Dionysios Filippas and Giorgos Dimitrakopoulos and Chrysostomos Nicopoulos},
  journal={Microprocess. Microsystems},
  year={2023},
  volume={102},
  pages={104938},
}

@article{Xue2025ReGateEP,
  title={ReGate: Enabling Power Gating in Neural Processing Units},
  author={Yu Xue and Jian Huang},
  journal={Proceedings of the 58th IEEE/ACM International Symposium on Microarchitecture},
  year={2025},
}

@inproceedings{sun2024simple,
  title={A simple and effective pruning approach for large language models},
  author={Sun, Mingjie and Liu, Zhuang and Bair, Anna and Kolter, Zico},
  booktitle={International Conference on Learning Representations},
  volume={2024},
  pages={4942--4964},
  year={2024}
}

@article{Frantar2023SparseGPTML,
  title={SparseGPT: Massive Language Models Can Be Accurately Pruned in One-Shot},
  author={Elias Frantar and Martin Jaggi},
  journal={ArXiv},
  year={2023},
  volume={abs/2301.00774},
}

@inproceedings{LeCun1989OptimalBD,
  title={Optimal Brain Damage},
  author={Yann LeCun and John S. Denker and Sara A. Solla},
  booktitle={Neural Information Processing Systems},
  year={1989},
}

@inproceedings{Wolf2013YosysAFV,
  title={Yosys-A Free Verilog Synthesis Suite},
  author={Clifford Wolf and Johann Glaser and Johannes Kepler},
  year={2013},
}

@inproceedings{Ajayi2019OpenROADTA,
  title={OpenROAD: Toward a Self-Driving, Open-Source Digital Layout Implementation Tool Chain},
  author={Tutu Ajayi and David Blaauw and Tuck-Boon Chan and C.-K. Cheng and Vidya A. Chhabria and D. K. Choo and M. Coltella and Ronald G. Dreslinski and Mateus Fogaça and S. Mehdi Hashemi and A. A. Ibrahim and Andrew B. Kahng and M. Kim and J. Li and Z. Liang and Uday Mallappa and Paul I. P{\'e}nzes and Geraldo Pradipta and Sherief Reda and Austin Rovinski and Kambiz Samadi and Sachin S. Sapatnekar and Lawrence K. Saul and Carl Sechen and Vaishnav Srinivas and William Swartz and Dennis Sylvester and D. Urquhart and L. Wang and Ming-Yu Woo and B. Xu},
  year={2019},
}

@article{Wu2019AccelergyAA,
  title={Accelergy: An Architecture-Level Energy Estimation Methodology for Accelerator Designs},
  author={Yannan Nellie Wu and Joel S. Emer and Vivienne Sze},
  journal={2019 IEEE/ACM International Conference on Computer-Aided Design (ICCAD)},
  year={2019},
  pages={1-8},
}

@misc{CherryOpenSTA,
  author       = {Cherry, James and Scott, William and  {Parallax Software, Inc.}},
  title        = {{OpenSTA}: Parallax Static Timing Analyzer},
  year         = {2026},
  publisher    = {GitHub},
  journal      = {GitHub repository},
  howpublished = {\url{https://github.com/The-OpenROAD-Project/OpenSTA}},
}

@misc{TILOS_NanGate45,
  author       = {{TILOS AI Institute}},
  title        = {{NanGate45 (FreePDK45, 45nm Open Cell Library, bsg\_fakeram memory generation): MacroPlacement}},
  year         = {2026},
  howpublished = {\url{https://tilos-ai-institute.github.io/MacroPlacement/Enablements/NanGate45/}},
  note         = {Accessed: 2026-11-16}
}

@article{Gen2019GemminiES,
  title={Gemmini: Enabling Systematic Deep-Learning Architecture Evaluation via Full-Stack Integration},
  author={Hasan Genç and Seah Kim and Alon Amid and Ameer Haj-Ali and Vighnesh Iyer and Pranav Raj Prakash and Jerry Zhao and Daniel Grubb and Harrison Liew and Howard Mao and Albert J. Ou and Colin Schmidt and Samuel Steffl and John Charles Wright and Ion Stoica and Jonathan Ragan-Kelley and Krste Asanovi{\'c} and Borivoje Nikoli{\'c} and Yakun Sophia Shao},
  journal={2021 58th ACM/IEEE Design Automation Conference (DAC)},
  year={2019},
  pages={769-774},
}

@article{Patel2023SplitwiseEG,
  title={Splitwise: Efficient Generative LLM Inference Using Phase Splitting},
  author={Pratyush Patel and Esha Choukse and Chao-Jie Zhang and Aashaka Shah and {\'I}{\~n}igo Goiri and Saeed Maleki and Ricardo Bianchini},
  journal={2024 ACM/IEEE 51st Annual International Symposium on Computer Architecture (ISCA)},
  year={2023},
  pages={118-132},
}

\end{document}